\PassOptionsToPackage{numbers,sort&compress}{natbib} 
\documentclass[%
preprint,
titlepage,
superscriptaddress,
nofootinbib,
 amsmath,amssymb,
 aps,
]{revtex4-2}

\usepackage[dvipdfmx]{graphicx}
\usepackage{subcaption}
\usepackage{dcolumn}
\usepackage{bm}
\usepackage{threeparttable}
\usepackage{caption}
\usepackage{xcolor}
\usepackage[colorlinks=true, citecolor=blue, linkcolor=blue, urlcolor=blue]{hyperref}
\usepackage[section]{placeins}
\begin{document}

\preprint{}

\title{
    Exponential Quintessence: Analytic Relationship Between the Current Equation of State Parameter and the Potential Parameter 
}
\author{Naoto Maki}
\affiliation{Department of Astronomical Science, The Graduate University for Advanced Studies (SOKENDAI), 2-21-1 Osawa, Mitaka, Tokyo 181-8588, Japan}
\affiliation{Division of Science, National Astronomical Observatory of Japan, 2-21-1 Osawa, Mitaka, Tokyo 181-8588, Japan}

\author{Kazunori Kohri}
\affiliation{Division of Science, National Astronomical Observatory of Japan, 2-21-1 Osawa, Mitaka, Tokyo 181-8588, Japan}
\affiliation{Department of Astronomy, The University of Tokyo, Bunkyo-ku, Hongo, Tokyo 113-0033, Japan}
\affiliation{Department of Astronomical Science, The Graduate University for Advanced Studies (SOKENDAI), 2-21-1 Osawa, Mitaka, Tokyo 181-8588, Japan}
\affiliation{Theory Center, IPNS, KEK,
1-1 Oho, Tsukuba, Ibaraki 305-0801, Japan}
\affiliation{Kavli IPMU (WPI), UTIAS, The University of Tokyo, Kashiwa, Chiba 277-8583, Japan}

\begin{abstract}
    Motivated by the indications of time-varying dark energy equation of state reported from DESI, we investigate a quintessence model with an exponential potential $V_0 e^{-\lambda\phi/m_{\mathrm{pl}}}$. We derive an analytical relationship between the current equation of state parameter for the quintessence field and the potential parameter $\lambda$ required to realize radiation and matter domination. Our results provide a useful analytical relation for inferring the potential parameter $\lambda$ from the observed current equation of state parameter. Furthermore, based on this framework, we provide a new approximate analytical upper bound on the potential parameter $\lambda$ for current accelerated expansion. Concretely, we obtain $\lambda<1.94$ as an approximate analytic estimate by adopting $\Omega_{\phi0}=0.685$.
\end{abstract}

\maketitle

\section{Introduction}

Dark energy is an unknown energy component required to explain the accelerated expansion of the universe. In the standard $\Lambda\mathrm{CDM}$ model, this expansion is realized by a positive cosmological constant with an equation of state parameter satisfying $w=-1$. While the $\Lambda\mathrm{CDM}$ model has been consistent with observations so far, recent BAO (Baryon Acoustic Oscillation) data have indicated deviations, suggesting possible hints of time-evolution in the equation of state parameter $w$~\cite{DESI:2024mwx,DESI:2025zgx}. (For general interpretation of the DESI results, see~\cite{Tada:2024znt,Yang:2024kdo,RoyChoudhury:2025iis,Maqsood:2026krg,Woo:2026ice,Colgain:2025nzf}.)

As a simple model which utilizes a canonical scalar field to drive accelerated expansion, the quintessence model is attracting attention again. (However, a canonical scalar field with $V(\phi)>0$ cannot create phantom behavior, which is suggested by DESI. Additional extensions or modifications such as dark-sector interactions, modified gravity, non-minimal couplings, or non-canonical kinetic terms are necessary to explain phantom behavior and phantom crossing \cite{Shlivko:2026jxa,Wolf:2025jed,Chanda:2026dmx,Chen:2025ywv,Granda:2010hb}. See Ref.~\cite{Li:2026asg} for the recent review.)
 Among various potential forms, the exponential potential $V(\phi)=V_0 e^{-\lambda\phi/m_{\mathrm{pl}}}$ has been studied extensively using dynamical systems analysis \cite{Halliwell:1986ja,Burd:1988ss,Ferreira:1997hj,Copeland:1997et,vandenHoogen:1999qq,Kolda:2001ex,Copeland:2006wr,Boehmer:2010jqg,Urena-Lopez:2011gxx,Tamanini:2014mpa,Gosenca:2015qha,Bahamonde:2017ize,SavasArapoglu:2017pyh,Andriot:2024jsh}. Here, $\lambda$ and $V_0$ are positive constants, and $m_{\mathrm{pl}}$ is the reduced Planck mass.
This exponential potential naturally appears in string theory~\cite{Obied:2018sgi,Nitta:2025mpi,Cicoli:2023opf}. It has also been studied in the context of modified gravity~\cite{Sadatian:2024oab,Hong:2025tyi,Devi:2026dwg} and quantum cosmology~\cite{Ratra:1989uv,Rebesh:2019pbw,Maki:2026fgm}. Multi-field extensions are discussed in Refs.~\cite{Liddle:1998jc,Malik:1998gy,Chiba:2014sda,Alestas:2025syk}. Observational constraints can be found in  Refs.~\cite{Bhattacharya:2024hep,Ramadan:2024kmn,Akrami:2025zlb,Bayat:2025xfr,Pourtsidou:2025sdd,BeltranJimenez:2026ymd,Sultana:2026ych,SanchezLopez:2025uzw}. Furthermore, it has been shown that utilizing the kination phase alleviates the fine-tuning problem of dark energy compared to the case of a cosmological constant \cite{Maki:2026qpf}. Extensions to interacting or non-minimally coupled dark energy models are discussed in \cite{Nozari:2016ilx,Pourtsidou:2025sdd,BeltranJimenez:2026ymd,Wang:2026gvg,Tzanni:2014eja,Wang:2026wrk}.

A significant recent development regarding exponential quintessence is the study of the correlation between the constant $\lambda$ and the current equation of state parameter of the scalar field $w_{\phi0}$ \cite{Andriot:2024jsh} (see also \cite{Andriot:2026lac,Andriot:2024sif}).
By imposing boundary conditions at the present epoch and fixing the density parameters, Ref.~\cite{Andriot:2024jsh} numerically demonstrated that for a given $\lambda$, the values of $w_{\phi0}$ that successfully realize radiation and matter domination are confined to a highly restricted region of the parameter space. Their numerical results indicate that the duration of radiation domination is extended as the degree of fine-tuning for $w_{\phi0}$ increases. Additionally, the study established an upper bound of $\lambda\lesssim 1.7683$ under the requirement of achieving both radiation and matter domination and current accelerated expansion.

In this study, we map out the parameter regions in the $(\lambda,w_{\phi0})$ space that yield solutions satisfying a diagnostic criterion adopted from Ref.~\cite{Andriot:2024jsh} and discuss the physical factors determining the shape of the distribution. Furthermore, for a given $\lambda$, we analytically derive the current equation of state parameter $w_{\phi0}$ required for a realistic solution in the form of a series expansion in terms of the scalar field density parameter $\Omega_\phi$. This allows us to determine the value of $w_{\phi0}$ for any given $\Omega_{\phi0}$ and $\lambda$. Therefore, our results provide an analytical explanation for the numerical finding in Ref.~\cite{Andriot:2024jsh}. Finally, using this analytical expression, we derive a universal relation that establishes the upper bound of $\lambda$ consistent with the current accelerated expansion of the universe.

In Sec.~\ref{sec:Formalism}, we formulate the cosmological equations for a system including quintessence in a flat, homogeneous, and isotropic universe with matter and radiation using a dynamical systems approach. Next, in Sec.~\ref{sec:Numerical_Analysis_of_the_Parameter_Space}, we numerically explore the $(\lambda,w_{\phi0})$ parameter space, uncovering the distribution of solutions that realize radiation and matter domination and the underlying mechanism governing its structure. Finally, in Sec.~\ref{sec:Analytical_Derivation}, we analytically derive the relation between $\lambda$ and $w_{\phi0}$ required for realistic solutions via a series expansion and demonstrate its consistency with the numerically obtained parameter distributions. Appendix~\ref{sec:Series_Expansion_of_Equation_of_State_Parameter around_Fixed_Point_C} provides an analytical discussion of the separatrix by expanding the equation of state parameter around one of the fixed points.

Throughout this paper, we adopt units where $c=\hbar=k_B=1$.

\section{Formalism}
\label{sec:Formalism}
We assume a spatially flat Friedmann-Lema\^itre-Robertson-Walker (FLRW) metric:
\begin{align}
  ds^2=-dt^2 +a(t)^2 \left[ dr^2+r^2(d\theta^2+\sin^2 \theta d\phi^2) \right],
\end{align}
where $a(t)$ is the scale factor, normalized to unity today. We consider the following three energy components: radiation, matter and a real scalar field $\phi$. The scalar field has a canonical kinetic term and an exponential potential,
\begin{align}
  V(\phi)=V_0 e^{-\lambda \phi/m_{\mathrm{pl}}},
\end{align}
where $\lambda$ and $V_0$ are positive constants and $m_{\mathrm{pl}}$ is the reduced Planck mass. The energy density $\rho_\phi$ and pressure $p_\phi$ for the scalar field are given by
\begin{align}
  &\rho_\phi=\frac{1}{2}\dot{\phi}^2+V(\phi),\\
  &p_\phi=\frac{1}{2}\dot{\phi}^2-V(\phi).
\end{align}
where the over-dot denotes the derivative of $\phi$ with respect to cosmic time $t$. The equation of state parameter is given by
\begin{align}
  w_\phi\equiv p_\phi/\rho_\phi =\frac{\frac{1}{2}\dot{\phi}^2-V(\phi)}{\frac{1}{2}\dot{\phi}^2+V(\phi)}.
\end{align} 
where $w_\phi$ satisfies $-1\leq w_\phi \leq1$. The Einstein equation and the equation of motion for the scalar field are given by 
\begin{align}
  &3m_{\mathrm{pl}}^2 H^2 = \frac{1}{2}\dot{\phi}^2+V(\phi)+\rho_m+\rho_r\label{Friedmann equation},\\
  &2m_{\mathrm{pl}}^2 \dot{H}=-\dot{\phi}^2-\rho_m-\frac{4}{3}\rho_r\label{acceleration equation},\\
  &\ddot{\phi}+3H\dot{\phi}+V_{,\phi}=0,\label{scalar field EOM}
\end{align}
respectively, where $V_{,\phi}=\frac{dV}{d\phi}$. Furthermore,
following the notation of \cite{Bhattacharya:2024hep} we define the dimensionless variables as follows:
\begin{align}
  \Omega_m\equiv\frac{\rho_m}{3m_{\mathrm{pl}}^2H^2},\qquad x\equiv\frac{\dot{\phi}}{\sqrt{6}m_{\mathrm{pl}} H},\qquad y\equiv\frac{\sqrt{V}}{\sqrt{3}m_{\mathrm{pl}} H},\qquad u\equiv\frac{\sqrt{\rho_r}}{\sqrt{3}m_{\mathrm{pl}}H}.
  \label{dimensionless parameter}
\end{align}
The squares of $x,y,u$ correspond to the density parameter of the
scalar field kinetic energy $\Omega_{\mathrm{kin}}\equiv x^2$, the
potential energy density parameter $\Omega_{\mathrm{pot}}\equiv y^2$,
and the radiation density parameter $\Omega_r\equiv u^2$,
respectively. Here, denoting the derivative with respect to the
e-folding number $N=\ln a$ by a prime, $'=\frac{d}{dN}$, we obtain the following
relations for $x,y$ and $u$:
\begin{align}
  &x'=\frac{1}{2}x(-3+u^2+3x^2-3y^2)+\sqrt{\frac{3}{2}}\lambda y^2,\label{eq:x}\\
  &y'=\frac{1}{2}y(3+u^2+3x^2-3y^2)-\sqrt{\frac{3}{2}}\lambda xy,\label{eq:y}\\
  &u'=\frac{1}{2}u(-1+u^2+3x^2-3y^2).\label{eq:u}
\end{align}
The fixed points of this system of equations are given in Table~\ref{tab:fixedpoint}.
\begin{table}[htbp]
\centering
\renewcommand{\arraystretch}{1.5} 
\resizebox{\textwidth}{!}{
\begin{tabular}{|c|c|c|c|c|} 
\hline

&Fixed points $(x,y,u)$ &Eigenvalues & Existence & Stability \\
\hline \hline

$A_{+}$ &$(+1,0,0)$ &  $\left( 3,3- \lambda\sqrt{\frac{3}{2}},1 \right)$ &$\forall\lambda$ & \begin{tabular}[c]{@{}c@{}}  Unstable for  $\lambda\le\sqrt{6}$ \\ \small Saddle for $\lambda>\sqrt{6}$\end{tabular} \\
\hline

$A_{-}$ &$(-1,0,0)$ &  $\left( 3,3+ \lambda\sqrt{\frac{3}{2}},1 \right)$ &$\forall\lambda$ & \begin{tabular}[c]{@{}c@{}}  Unstable 
\end{tabular} \\
\hline

$B$ &$(0,0,0)$ & $\left( -\frac{3}{2},\frac{3}{2},-\frac{1}{2} \right)$ &  $\forall\lambda$ & Saddle \\
\hline

$C$ &$\left( \frac{\lambda}{\sqrt{6}},\pm\frac{\sqrt{6-\lambda^{2}}}{\sqrt{6}},0 \right)$ & $\left( \frac{\lambda^2}{2}-3,\lambda^2-3,\frac{\lambda^2}{2}-2 \right)$ & $\lambda<\sqrt{6}$  & \begin{tabular}[c]{@{}c@{}}Stable for $\lambda<\sqrt{3}$\\ Saddle for $\lambda>\sqrt{3}$\end{tabular} \\
\hline

$D$ &$\left( \frac{1}{\lambda}\sqrt{\frac{3}{2}},\pm\frac{1}{\lambda}\sqrt{\frac{3}{2}},0 \right)$ &  $\left( -\frac{3(\lambda+\sqrt{24-7\lambda^2})}{4\lambda}, -\frac{3(\lambda-\sqrt{24-7\lambda^2})}{4\lambda} ,-\frac{1}{2}\right)$& $\lambda>\sqrt{3}$ & Stable \\
\hline

$E$ &$(0,0,\pm1)$ & $(-1,2,1)$ &$\forall\lambda$ & Saddle \\
\hline

$F$ &$\left( \frac{1}{\lambda}\sqrt{\frac{8}{3}},\pm\frac{2}{\lambda\sqrt{3}},\pm\sqrt{1-\frac{4}{\lambda^{2}}} \right)$  &$\left( -\frac{\lambda+\sqrt{64-15\lambda^2}}{2\lambda},-\frac{\lambda-\sqrt{64-15\lambda^2}}{2\lambda},1\right)$& $\lambda>2$ & Saddle \\
\hline
\end{tabular}
}
\caption{Each row displays the fixed point, the eigenvalues of the Jacobian, the condition of $\lambda$ for the existence, and the stability. We include unphysical fixed points with the negative values of $y$ and $u$. Note that we consider the cases only for $\Omega_k=0$. For cases including spatial curvature, see~\cite{Andriot:2024jsh,Bhattacharya:2024hep}.}
\label{tab:fixedpoint}
\end{table}
The points $A$, $B$, and $E$ correspond
to epochs dominated by the kinetic energy of the scalar field, matter, and radiation, respectively. $D$ and $F$ correspond to
the scaling solutions, in which the scalar field mimics the evolution of other energy components (i.e., matter and radiation). $C$ corresponds to an epoch dominated by the scalar field where its kinetic energy and potential energy evolve similarly.  The asymptotic state of the Universe depends on $\lambda$. When $\lambda < \sqrt{3}$, the Universe approaches the stable point $C$ where $w_\phi=w_{\mathrm{eff}}=-1+\frac{\lambda^2}{3}$. Conversely, when $\lambda \ge \sqrt{3}$, the point $D$ is stable where $w_\phi=w_{\mathrm{eff}}=0$. See Refs.~\cite{Copeland:2006wr,Andriot:2024jsh} for a review of the autonomous system and its fixed points.

\section{Numerical Analysis of the Parameter Space}
\label{sec:Numerical_Analysis_of_the_Parameter_Space}
In this section, we numerically explore the $(\lambda, w_{\phi 0})$ parameter space, where $\lambda$ is constant and $w_{\phi0}$ is the current equation of state parameter for the scalar field by imposing boundary conditions at the present epoch. 
As a diagnostic criterion for mapping solutions with expansion histories similar to our own universe, we select solutions that experience at least one period of radiation domination, defined by $\Omega_r > 0.5$, as in Ref.~\cite{Andriot:2024jsh}.
\footnote{Strictly speaking, radiation domination should persist until the epoch of Big Bang Nucleosynthesis ($T\sim \mathrm{MeV}$) in order not to change the abundance of light elements~\cite{Kawasaki:1999na,Kawasaki:2000en,Ichikawa:2005vw,deSalas:2015glj,Hasegawa:2019jsa,Barbieri:2025moq,Maki:2026qpf}. However, requiring $\Omega_r>0.5$ at least once provides a sufficient criterion for roughly mapping out the regions of the parameter space.}

\begin{figure}[htbp]
  \centering
  \includegraphics[scale = 0.7]{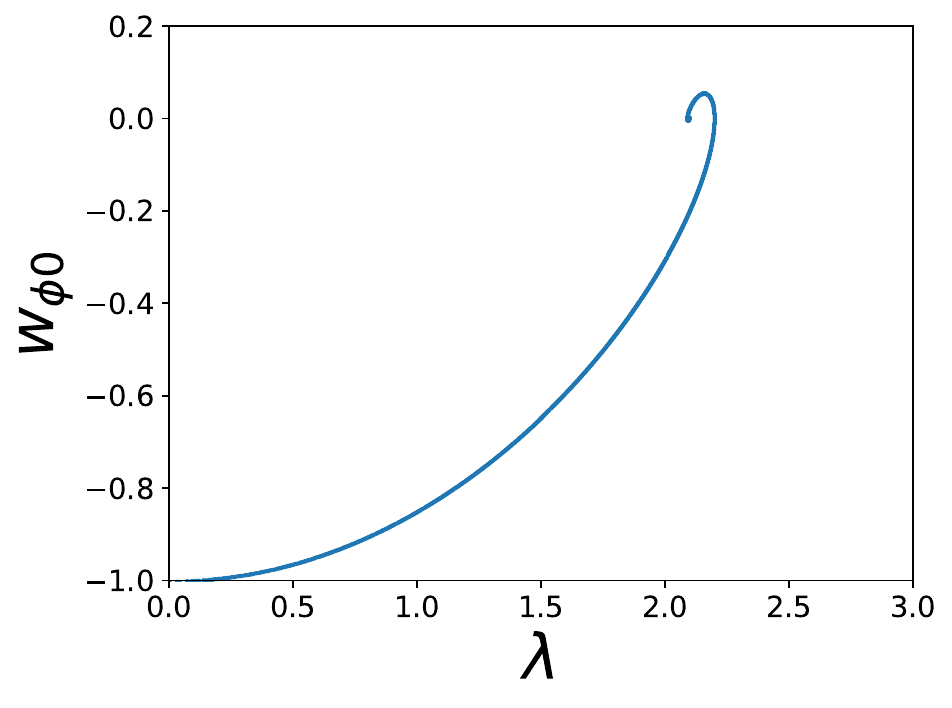}
  \caption{Scatter plot of the $(\lambda,w_{\phi0})$ pairs that yield solutions satisfying $\Omega_r>0.5$ at least once. The parameter scan was performed on a grid of $10^5\times10^4$ points in the $(\lambda,w_{\phi0})$ plane.}
  \label{fig:lambda_wphi0}
\end{figure}

\begin{figure}[htbp]
  \centering
  \begin{minipage}{.45\linewidth}
    \centering
    \includegraphics[width=\linewidth]{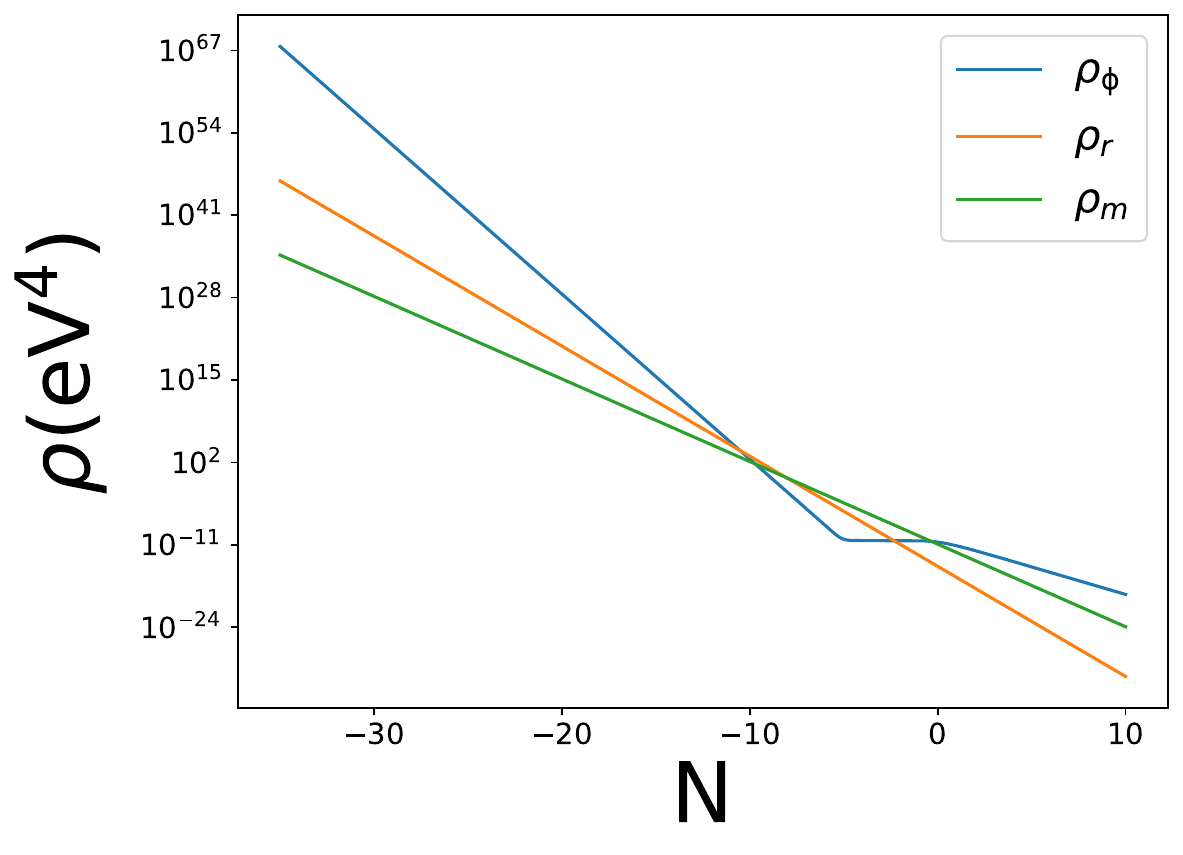}
    \subcaption{}
  \end{minipage}
  \hspace{1mm}
  \begin{minipage}{.45\linewidth}
    \centering
    \includegraphics[width=\linewidth]{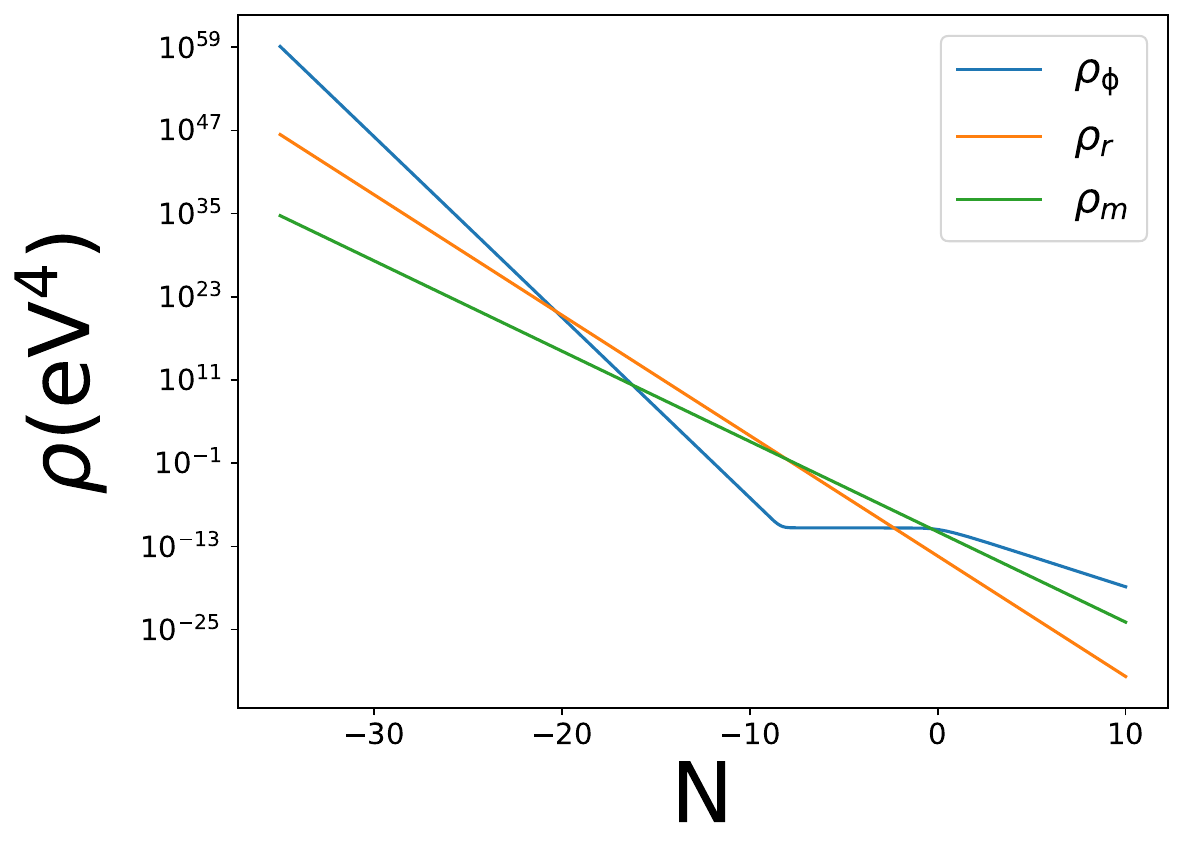}
    \subcaption{}
  \end{minipage}
  \vspace{3mm}
  \begin{minipage}{.45\linewidth}
    \centering
    \includegraphics[width=\linewidth]{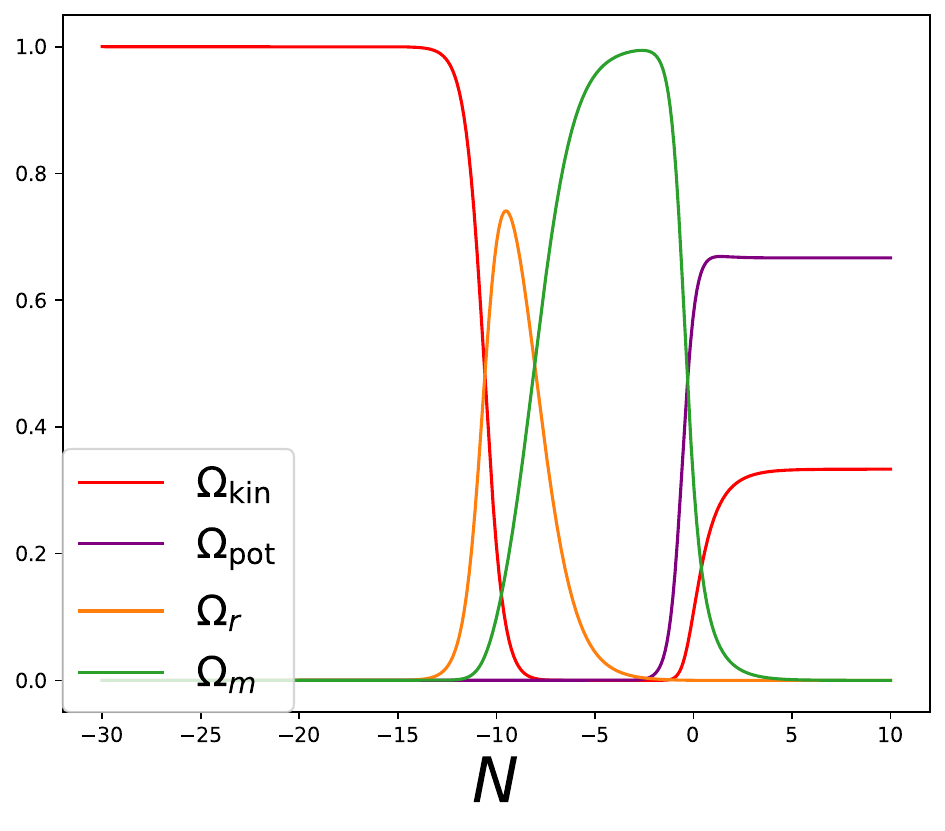}
    \subcaption{}
  \end{minipage}
  \hspace{1mm}
  \begin{minipage}{.45\linewidth}
    \centering
    \includegraphics[width=\linewidth]{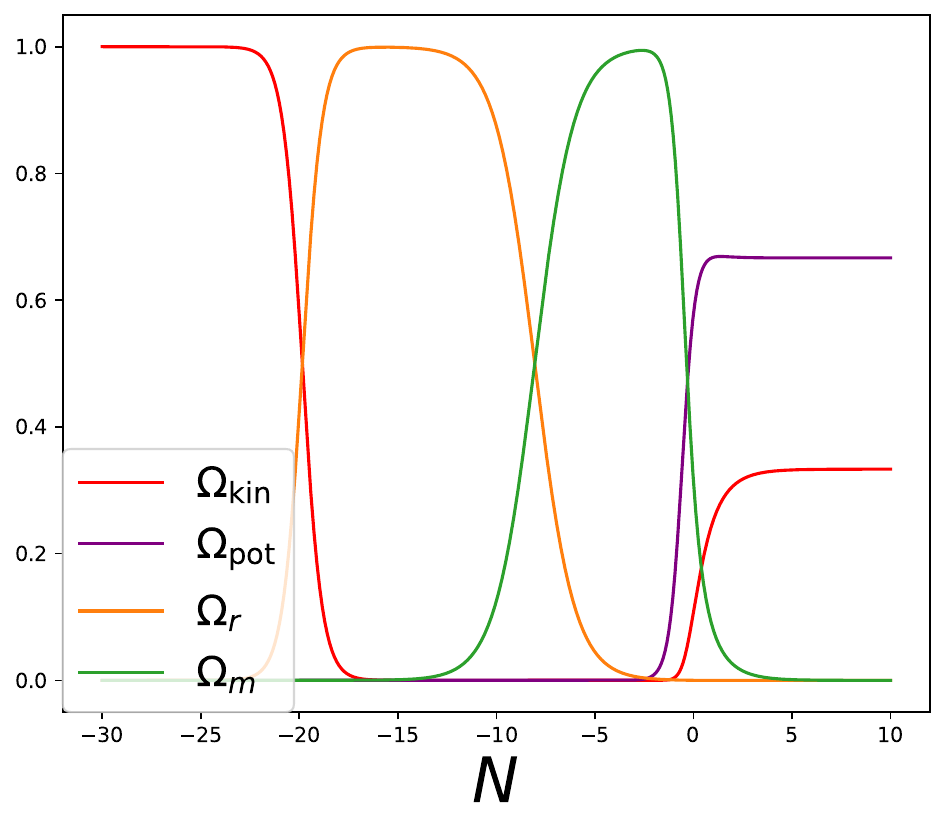}
    \subcaption{}
  \end{minipage}
  \caption{Evolution of the energy densities (a,b) and density parameters (c,d) for $\lambda=\sqrt{2}$. The left panels (a,c) correspond to $w_{\phi0}=-0.687473$, while the right panel (b,d) show the results for a more fine-tuned value, $w_{\phi0}=-0.687473593$. In the upper panels, blue, orange and green lines represent the energy densities of scalar field, radiation and matter, respectively. In the lower panels, the red, purple, orange, and green lines denote the density parameters of the scalar field kinetic energy ($\Omega_{\mathrm{kin}}= x^2$), potential energy $\Omega_{\mathrm{pot}}=y^2$, radiation ($\Omega_r=u^2$), and matter ($\Omega_m=1-x^2-y^2-u^2$), respectively. }
  \label{fig:rho_evolution}
\end{figure}

We solve the evolution of the relative energy densities based on Eqs.~\eqref{eq:x}--\eqref{eq:u}. Therefore, the required initial conditions are the values of $\Omega_{\phi0},\Omega_{m0},\Omega_{r0},w_{\phi0}$ and the constant $\lambda$ (note that one density parameter is not independent due to the constraint $1=\Omega_{\phi0}+\Omega_{m0}+\Omega_{r0}$). In this setup, providing $w_{\phi0}$ is equivalent to specifying the ratio between the kinetic term $x$ and the potential term $y$ of the current energy density of the scalar field. Here, we assume that the current $\dot{\phi}$ is positive. \enlargethispage{-2\baselineskip}Adopting the observational data~\cite{Planck:2018vyg}, we use the following parameter set (see also~\cite{Andriot:2024jsh}):\footnote{While $\Omega_\phi$ should be determined through a new MCMC analysis for the present model, we adopt the value for the $\Lambda$CDM model because the precise choice of $\Omega_\phi$ does not essentially affect the following discussion. We use these values whenever the current density parameters are required.}
\begin{align}
  \Omega_{\phi0}=0.6850,\quad\Omega_{m0}=0.3149,\quad\Omega_{r0}=0.0001.
  \label{eq:boundary_condition}
\end{align}
Note that the analytical results presented later are independent of the specific values of current density parameters chosen here. Fig.~\ref{fig:lambda_wphi0} shows a scatter plot of the $(\lambda,w_{\phi0})$ pairs that yield solutions satisfying $\Omega_r >0.5$ at least once.  
It can be seen that once $\lambda$ is fixed, the value of $w_{\phi0}$ which realizes a radiation-dominated era is almost uniquely determined. Furthermore, the figure indicates the existence of an upper bound for $\lambda$. Fig.~\ref{fig:rho_evolution} shows the evolution of energy densities and density parameters for different $w_{\phi0}$ values with the same $\lambda=\sqrt{2}$. 
As can be seen from Fig.~\ref{fig:rho_evolution}, the kinetic energy of the scalar field dominates the early universe. This state is known as kination. During the kination phase, the kinetic energy of the scalar field evolves as $\dot{\phi}^2/2\propto a^{-6}$. Increasing the degree of fine-tuning for $w_{\phi0}$ corresponds to extending the period where the energy density of the scalar field is nearly constant. Prolonging the phase where the scalar field behaves like a cosmological constant shifts the transition between kination and radiation domination to an earlier time, consequently expanding the total duration of the radiation-dominated era.

\begin{figure}[htbp]
  \centering
  \includegraphics[scale = 0.65]{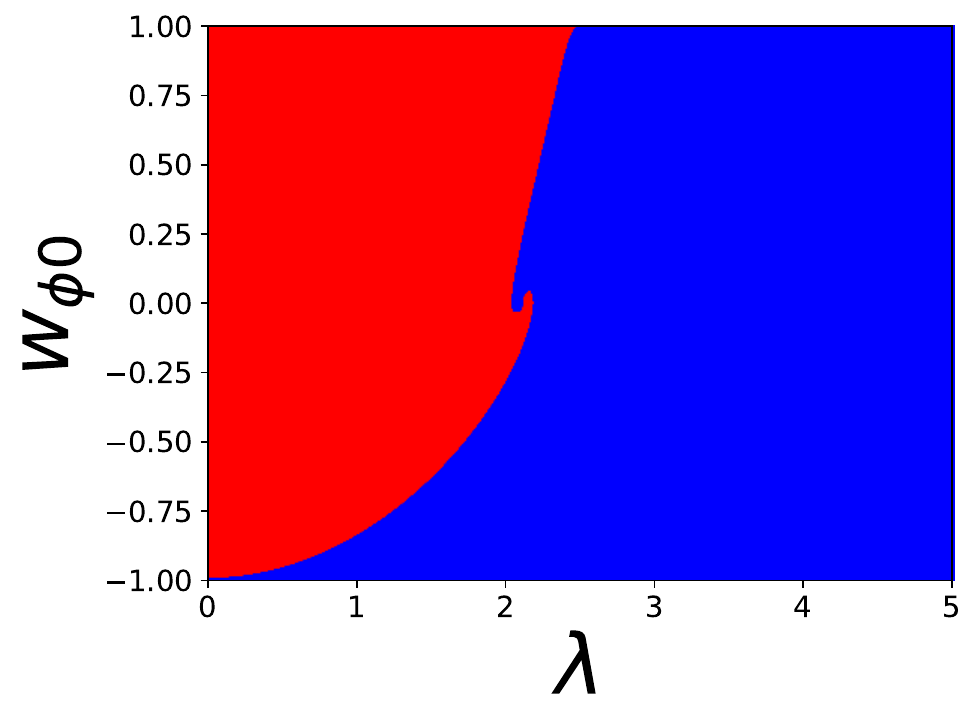}
  \caption{Scatter plot of the sign of $\dot{\phi}$ at $N=-30$ in the $(\lambda,w_{\phi0})$ plane. The system is integrated backward from $N=0$ to $N=-30$ with the boundary condition at present time in Eq.~\ref{eq:boundary_condition}. Red points (region) correspond to the case where $\dot{\phi}>0$ at $N=-30$, representing solutions originating from the $A_+$. Conversely, blue points (region) correspond to the case where $\dot{\phi}<0$ at $N=-30$, representing solutions originating from the $A_-$.The parameter scan was performed on a grid of $5000\times5000$ points in the $(\lambda,w_{\phi0})$ plane.}
  \label{fig:separatrix}
\end{figure}

Next, we discuss the physical origin of the distribution of parameters $(\lambda,w_{\phi0})$ in Fig.~\ref{fig:lambda_wphi0} that realize radiation domination. The distribution observed in Fig.~\ref{fig:lambda_wphi0} is essentially governed by the two repellers in this dynamical system, denoted as points $A_+$ and $A_{-}$ in Table~\ref{tab:fixedpoint}. In this context, a repeller is defined as a fixed point for which all eigenvalues of the Jacobian have positive real parts. A repeller is essentially the opposite of an attractor, acting as a source from which trajectories originate in phase space.
These repellers, $A_+$ and $A_{-}$, correspond to kination phases with $\dot{\phi}>0$ and $\dot{\phi}<0$, respectively. When solving the system backward in time, trajectories converge toward these repellers. 
Fig.~\ref{fig:separatrix} presents a scatter plot where the parameter space is color-coded in red and blue according to whether they converge toward the repeller $A_+$ or $A_-$ when evolved backward in time from the present epoch.
For $w_{\phi0}\lesssim0$, the boundary in the ($\lambda,w_{\phi0}$) parameter space separating the regions that converge toward $A_+$ and $A_-$ coincides with the distribution shown in Fig.~\ref{fig:lambda_wphi0}. 
This implies that solutions satisfying our diagnostic criterion are located on the separatrix between these two repellers. To achieve a radiation-dominated era, it is necessary to prolong the period during which the scalar field behaves like a cosmological constant. This requires the backward-time trajectory to avoid falling into either $A_+$ or $A_-$ for as long as possible. Consequently, the parameter region that avoids immediate convergence toward either repeller is exactly the one that yields solutions satisfying our diagnostic criterion, as shown in Fig.~\ref{fig:lambda_wphi0}. While we have assumed $\dot{\phi}>0$ at present, it is worth discussing the case of $\dot{\phi}<0$. If we assume the boundary condition $\dot{\phi}<0$ today, the trajectories for all pairs of $(\lambda,w_{\phi 0})$ immediately converge to the fixed point $A_{-}$ when integrated backward in time. This means the separatrix does not exist in this situation. Consequently, the scalar field cannot experience a phase where it behaves like a cosmological constant for a sufficient duration in the past. This makes it impossible to construct cosmological solutions that realize a radiation-dominated era and a proper duration of matter domination.

It should be noted, however, that there exists another separatrix at $w_{\phi0}\gtrsim0$ that has no counterpart in Fig.~\ref{fig:lambda_wphi0}. These solutions approach the unstable fixed point $C$ in Table~\ref{tab:fixedpoint} where $\Omega_\phi=1$ and the ratio of the kinetic energy and potential energy is constant ($x^2=\frac{\lambda}{6},y^2=1-\frac{\lambda^2}{6}$). Since these solutions fail to realize matter or radiation domination eras, they do not appear in Fig.~\ref{fig:lambda_wphi0}. We discuss this separatrix analytically in Appendix~\ref{sec:Series_Expansion_of_Equation_of_State_Parameter around_Fixed_Point_C}.

These separatrices exist only for $\lambda<\sqrt{6}$. This is because, for $\lambda>\sqrt{6}$, $A_+$ loses its character as a repeller as one of the eigenvalues of the Jacobian at $A_+$ becomes negative. As a result, the separatrix disappears for $\lambda > \sqrt{6}$, and all trajectories that reach the present epoch originate from the $\dot{\phi}<0$ region when solved backward in time.

\begin{figure}[htbp]
  \centering
  \includegraphics[scale = 0.8]{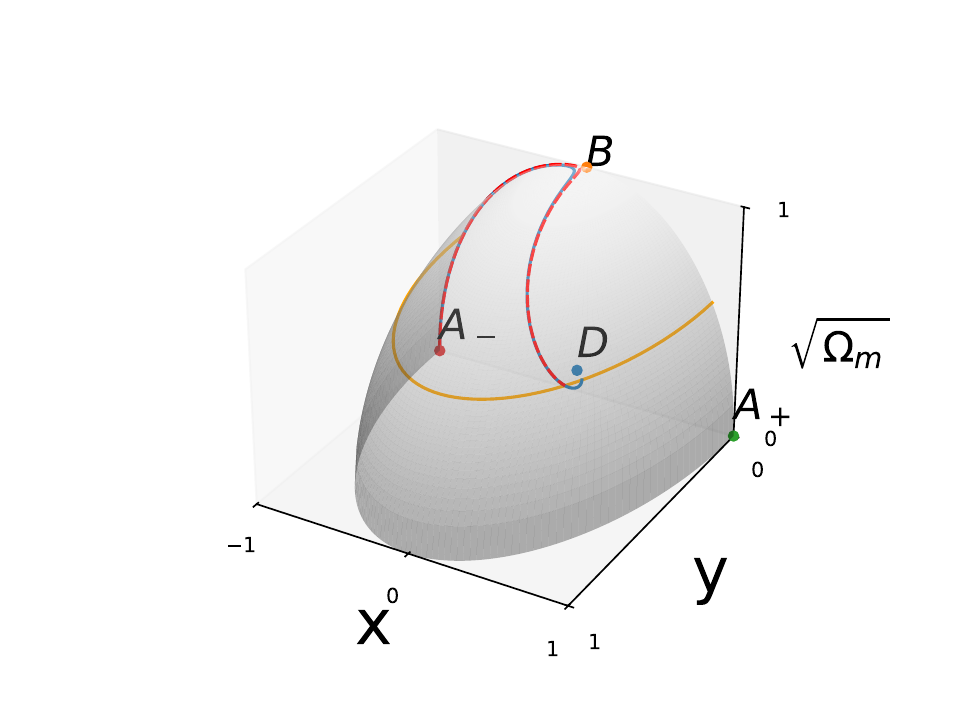}
  \caption{
  Comparison of two trajectories satisfying our diagnostic criterion for $\lambda=2.155$ in the phase space $(x,y,\sqrt{\Omega_m})$. The red dashed line ($w_{\phi0}=0.053782$) and the blue solid line ($w_{\phi0}=-0.117749$) represent two possible cosmic histories for the same potential parameter. The orange line represents the current density parameter $\sqrt{\Omega_{m0}}$. The choice of phase space coordinates follows \cite{Andriot:2024jsh}. }
  \label{fig:double-valued}
\end{figure}

We now discuss the origin of two branches of $w_{\phi0}$ observed around $\lambda\sim 2.1$ in Fig.~\ref{fig:lambda_wphi0}, where a single value of $\lambda$ can yield two distinct solutions satisfying our diagnostic criterion.
As an example, Fig.~\ref{fig:double-valued} plots the trajectories in the phase space $(x,y,\sqrt{\Omega_m})$ for these two solutions at $\lambda=2.155$, following the visualization approach in \cite{Andriot:2024jsh}. 
This figure reveals that the two solutions actually correspond to a single trajectory in the phase space. Since this trajectory passes through the current value $\sqrt{\Omega_m}=\sqrt{\Omega_{m0}}$ at two distinct points, two different choices for $w_{\phi0}$ exist depending on which of these points is identified as the present state. This behavior occurs because the eigenvalues of fixed point $D$ become complex for $\lambda>\sqrt{24/7}$, which induces a spiral approach toward the fixed point in the phase space.

\section{Analytical Derivation}
\label{sec:Analytical_Derivation}
\begin{figure}[htbp]
  \centering
  \includegraphics[scale = 0.8]{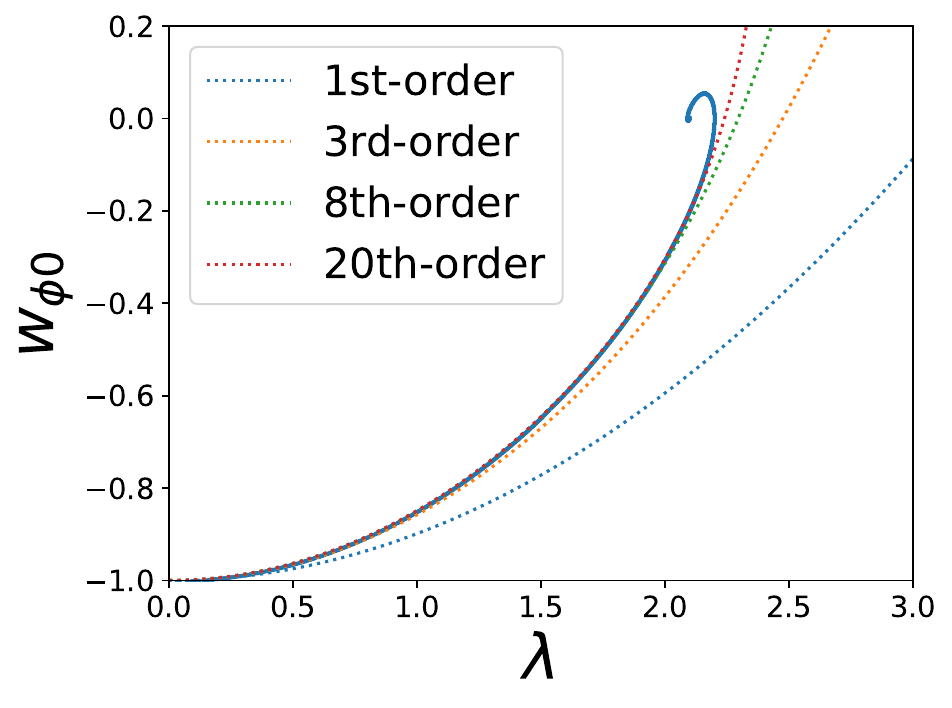}
  \caption{Numerical results and analytical prediction for the pairs of $(\lambda,w_{\phi0})$ that realize a radiation-dominated era. The blue dots represent the numerical results in Fig.~\ref{fig:lambda_wphi0} obtained through a parameter scan of the $(\lambda,w_{\phi0})$ pairs that yield solutions satisfying $\Omega_r>0.5$ at least once. The dotted lines correspond to the theoretical prediction based on the series expansion of $\Omega_\phi$ up to the 1st, 3rd, 8th and 20th orders. }
  \label{fig:w_phi_analytic}
\end{figure}

In this section, we analytically derive the sets of $(\lambda,w_{\phi0})$ that yield solutions satisfying the criterion, which were numerically determined in Sec.~\ref{sec:Numerical_Analysis_of_the_Parameter_Space}. Since the energy density of radiation is negligible in the vicinity of the present time, we can assume $u\ll x,y<1$. By neglecting $u$ in Eqs.~\eqref{eq:x}--\eqref{eq:u}, the equations reduce to:
\begin{align}
  &x'=\frac{1}{2}x(-3+3x^2-3y^2)+\sqrt{\frac{3}{2}}\lambda y^2,\\
  &y'=\frac{1}{2}y(3+3x^2-3y^2)-\sqrt{\frac{3}{2}}\lambda xy.\\
\end{align}
By differentiating $w_{\phi}=(x^2-y^2)/(x^2+y^2)$ and $\Omega_\phi=x^2+y^2$ with respect to the e-folding number $N$, we obtain the following relations:
\begin{align}
    \frac{dw_\phi}{dN}&=-(1-w_\phi)\left\{3(1+w_\phi) -\lambda\sqrt{3\Omega_\phi(1+w_\phi)} \right\},\\
    \frac{d\Omega_\phi}{dN}&=-3w_\phi \Omega_\phi(1-\Omega_\phi).
\end{align}
From these equations, we obtain:
\begin{align}
    \frac{dw_\phi}{d\Omega_\phi}=\frac{(1-w_\phi)\left\{ 3(1+w_\phi)-\lambda\sqrt{3\Omega_\phi(1+w_\phi)} \right\}}{3w_\phi\Omega_\phi(1-\Omega_\phi)}.
    \label{dwdOmega}
\end{align}
Here, solutions that realize sufficiently long matter and radiation domination must satisfy $w_\phi\to -1$ and $\Omega_\phi \to0 $ in the backward evolution from the present (i.e., they should deviate from $w_\phi=-1$ and increase in $\Omega_\phi$ only in the vicinity of the present epoch). Therefore, we expand the deviation of $w_\phi$ from $-1$, namely $1+w_\phi$, as a power series in $\Omega_\phi$:
\begin{align}
    1+w_\phi=\sum_{n=1}^{\infty}c_n \Omega_\phi^n,
\end{align}
where $c_n$ are constants only depending on $\lambda$. In the following, as an example, we determine the coefficients $c_n$ up to the third order in $\Omega_\phi$. Substituting the expansion 
\begin{align}
   1+ w_\phi\simeq  c_1\Omega_\phi +c_2\Omega_\phi^2+c_3\Omega_\phi^3,
\end{align}
into Eq.~\eqref{dwdOmega} and comparing the coefficients for each order of $\Omega_\phi$, we obtain the following relations for the first, second, and third order terms:
\begin{align}
    -3c_1&=2(3c_1-\lambda\sqrt{3c_1}),\\
    -3(2c_2-c_1) +3c_1^2 &=2\left( 3c_2-\lambda\sqrt{3c_1}\frac{c_2}{2c_1} \right)-c_1\left( 3c_1 -\lambda\sqrt{3c_1} \right),\\
    -3(3c_3-2c_2)+3c_1 (2c_2-c_1)+3c_1c_2 &=2\left( 3c_3 -\lambda\sqrt{3c_1}\left( \frac{c_3}{2c_1}-\frac{c_2^2}{8c_1^2} \right)\right)\notag\\
    &\qquad-c_1\left( 3c_2-\lambda\sqrt{3c_1}\frac{c_2}{2c_1} \right)-c_2(3c_1-\lambda\sqrt{3c_1}).
\end{align}
From the first equation, we obtain:
\begin{align}
    c_1=\frac{4}{27}\lambda^2.
\end{align}
Substituting this into the second equation yields:
\begin{align}
    c_2=\frac{16}{3645}\lambda^4+\frac{8}{135}\lambda^2.
\end{align}
Finally, substituting $c_1$ and $c_2$ into the third equation, we find:
\begin{align}
    c_3&=\frac{1712}{3444525}\lambda^6+\frac{352}{127575}\lambda^4+\frac{148}{4725}\lambda^2.
\end{align}
Consequently, we obtain the series expansion of $w_{\phi0}$ up to the third order in $\Omega_{\phi0}$:
\begin{align}
    w_{\phi0}\simeq-1+\frac{4}{27}\lambda^2\Omega_{\phi0}+\left(\frac{16}{3645}\lambda^4+ \frac{8}{135}\lambda^2 \right)\Omega_{\phi0}^2+\left( \frac{1712}{3444525}\lambda^6+\frac{352}{127575}\lambda^4+\frac{148}{4725}\lambda^2 \right) \Omega_{\phi0}^3.
    \label{eq:expansion_of_w}
\end{align}
This takes the form $w_{\phi0}=w_{\phi0}(\lambda)$ and can be regarded as a function that returns the values of $w_{\phi0}$ yielding a solution on the separatrix for a given $\lambda$. In Fig.~\ref{fig:w_phi_analytic}, the analytical predictions are compared against the numerical result in Fig.~\ref{fig:lambda_wphi0}. 
 It is evident that the accuracy improves as the order of the expansion increases. For instance, when $\lambda=1$, the expansion up to the first order yields a relative error of $5.9\%$ compared to the true value of $w_{\phi0}$. This error reduces to $2.4\%$ and $1.0\%$ when including terms up to the second and third orders, respectively. Based on the empirical comparison in Fig.~\ref{fig:w_phi_analytic}, we propose $\lambda \lesssim 2$ as the practical domain of validity for this analytical expansion. This range is practically sufficient as it fully encompasses the parameter region where the current accelerated expansion is realized. Because our series expansion is inherently single-valued, it is intended only for the lower branch ($w_{\phi0}<0$).
By retaining only the lowest-order term of $\gamma\equiv 1+w_\phi$ in Eq.~\eqref{dwdOmega}, Ref.~\cite{Scherrer:2007pu} analytically solved it. Their approximate solution corresponds to the terms proportional to $\lambda^2$ in Eq.~\eqref{eq:expansion_of_w}. In our derivation, by taking a different approach, namely, expanding $1+w_\phi$ in terms of $\Omega_\phi$, we derived higher-order terms in $\lambda$, including $\lambda^4$ and $\lambda^6$, which were neglected in their work. Therefore, our work can be regarded as an extension of Ref.~\cite{Scherrer:2007pu} to higher orders in $\lambda$.

While the above results analytically reproduce the values of $w_{\phi0}$ that yield a radiation-dominated era, the region of the $(\lambda,w_{\phi0})$ parameter space realizing this era corresponds only to the lower half ($w_{\phi0}<0$) of the separatrix between $A_{-}$ and $A_{+}$. The upper half of the separatrix ($w_{\phi0}>0$) can be obtained by performing an expansion around $-1+\frac{\lambda^2}{3}$ in terms of $1-\Omega_\phi$. We provide a detailed derivation in the Appendix~\ref{sec:Series_Expansion_of_Equation_of_State_Parameter around_Fixed_Point_C}.

Furthermore, the condition for current accelerated expansion, $w_{\phi0}<-\frac{1}{3\Omega_{\phi0}}$, can be expressed using the first-order approximation of $w_{\phi0}$ in terms of $\Omega_{\phi0}$ as follows:
\begin{align}
  \lambda <\frac{3\sqrt{3\Omega_{\phi0}-1}}{2\Omega_{\phi0}}.
  \label{eq:lambda_upper_bound_first_order}
\end{align}
For $\Omega_{\phi0}=0.685$, this yields $\lambda<2.25$. By using the second-order approximation of $w_{\phi0}$ in terms of $\Omega_{\phi0}$, we obtain the following inequality:
\begin{align}
  \lambda<\sqrt{\frac{-27\sqrt{\Omega_{\phi0}}(2\Omega_{\phi0}+5)+9\sqrt{3(12\Omega_{\phi0}^3+60\Omega_{\phi0}^2+135\Omega_{\phi0}-20)}}{8\Omega_{\phi0}^{3/2}}}.
  \label{eq:lambda_upper_bound_second_order}
\end{align}
For $\Omega_{\phi0}=0.685$, this yields $\lambda<1.94$. Although these are looser bounds than the numerical constraint $\lambda<1.7683$ found in \cite{Andriot:2024jsh}, they provide approximate analytical estimates applicable to arbitrary $\Omega_{\phi0}$. Moreover, one can obtain tighter bounds by extending the expansion of $w_{\phi0}$ to higher orders as needed. We note, however, that the numerical approximate bound we obtained here assumes $\Omega_{\phi0}$ from the $\Lambda\mathrm{CDM}$ model. To derive a truly observationally consistent bound on $\lambda$, one must use the $\Omega_{\phi0}$ obtained by directly fitting this model to the data, which is beyond the scope of this paper.

\section{Discussion and Conclusion}
We have analyzed a quintessence model with an exponential potential $V=V_0 e^{-\lambda\phi/m_{\mathrm{pl}}}$, focusing on the relationship between the parameter $\lambda$ and the present equation of state parameter $w_{\phi0}$ when solved with boundary conditions set at the present epoch. Consistent with \cite{Andriot:2024jsh}, we find that solutions satisfying our diagnostic criterion for a radiation-dominated era only exist within a highly constrained range of $w_{\phi0}$ for a given $\lambda$ by using a numerical scan of $(\lambda,w_{\phi0})$ parameter space. To explain this result, we investigate the evolution of energy density and showed that to extend the radiation-dominated era, one must prolong the stage where the scalar field mimics a cosmological constant, which requires the fine-tuned boundary condition at the present time. Furthermore, we revealed that in the $(\lambda,w_{\phi0})$ parameter space, the region that realizes radiation domination exactly coincides with the separatrix between trajectories with initial $\dot{\phi}>0$ and those with $\dot{\phi}<0$.
Moreover, we derived an analytical expression for $w_{\phi0}$ as an infinite series of $\Omega_{\phi0}$ for a given $\lambda$. The expansion up to the third order in $\Omega_{\phi0}$ is:
\begin{align}
    w_{\phi0}\simeq-1+\frac{4}{27}\lambda^2\Omega_{\phi0}+\left(\frac{16}{3645}\lambda^4+ \frac{8}{135}\lambda^2 \right)\Omega_{\phi0}^2+\left( \frac{1712}{3444525}\lambda^6+\frac{352}{127575}\lambda^4+\frac{148}{4725}\lambda^2 \right) \Omega_{\phi0}^3.\notag
\end{align}
This result will be highly useful for inferring the potential parameter $\lambda$ from the observed current equation of state parameter $w_{\phi0}$.
Additionally, using the first-order and second-order expansions in $\Omega_{\phi0}$, we derived the approximate analytic conditions on $\lambda$ for the current accelerated expansion.
For $\Omega_{\phi0}=0.685$, the first-order approximation of $w_{\phi0}$ in terms of $\Omega_{\phi0}$ leads to $\lambda<2.25$. Furthermore, as demonstrated in Eq.~\eqref{eq:lambda_upper_bound_second_order}, using the second-order approximation of $w_{\phi0}$ provides a tighter approximate analytical bound, yielding $\lambda<1.94$.
While these represent relatively rougher approximations than the numerical upper bound found in Ref.~\cite{Andriot:2024jsh}, our analytical results provide approximate relationships applicable to any value of $\Omega_{\phi0}$. It is worth noting that the numerical approximate bounds on $\lambda$ we obtained rely on $\Omega_{\phi0}$ from the $\Lambda\mathrm{CDM}$ model. While direct data fitting using this model is necessary to obtain a truly consistent observational bound on $\lambda$, it is beyond the scope of this paper.

\begin{acknowledgments}
This work was in part supported by JSPS KAKENHI Grants
No. JP24K07027 (KK).
\end{acknowledgments}
\appendix

\section{Series Expansion of Equation of State Parameter Around Fixed Point C}
\label{sec:Series_Expansion_of_Equation_of_State_Parameter around_Fixed_Point_C}

\begin{figure}[htbp]
  \centering
  \includegraphics[scale = 0.6]{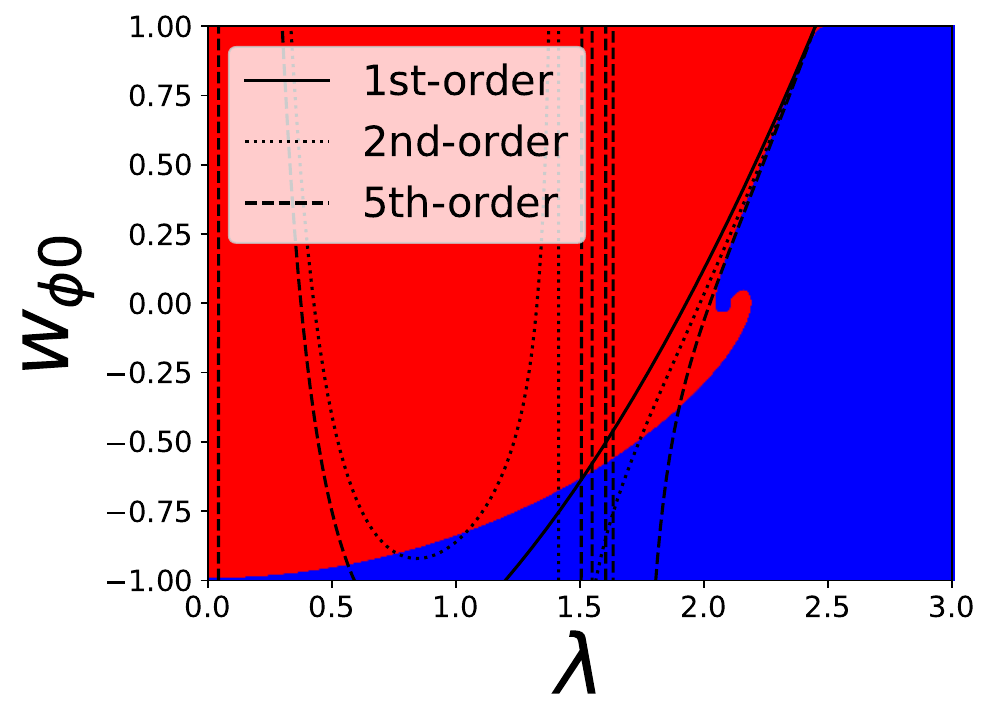}
  \caption{Comparison between the sign of $\dot{\phi}$ at $N=-30$ (red: $\dot{\phi}>0$, blue:$\dot{\phi}<0$) in the $(\lambda,w_{\phi0})$ parameter space and the theoretical curve derived by expanding $w_{\phi}$ around the fixed point $C$. Solid, dotted, and dashed lines represent the 1st, 2nd, and 5th order expansions, respectively.}
  \label{fig:separatrix_analytical}
\end{figure}

In this section, we derive the expansion of $w_\phi$ in terms of $\Omega_\phi$ around fixed point C. At this point, where $w_{\phi}=-1+\frac{\lambda^2}{3}$ and $\Omega_\phi=1$, we expand $w_\phi$ as follows:
\begin{align}
    w_\phi=-1+\frac{\lambda^2}{3}+\sum_{n=1}^{\infty}c_n (1-\Omega_\phi)^n.
\end{align}
Substituting this into Eq.~\ref{dwdOmega}, we obtain the expansion up to the second order:
\begin{align}
    w_\phi=-1+\frac{\lambda^2}{3}+\left( \frac{\lambda^2}{3}-2 \right)(1-\Omega_\phi)+\frac{-\lambda^6+8\lambda^4-15\lambda^2+18}{3\lambda^2(2-\lambda^2)}(1-\Omega_\phi)^2+\mathcal{O}\left( (1-\Omega_\phi)^3 \right).
\end{align}
As shown in Fig.~\ref{fig:separatrix_analytical}, the theoretical curve fits the separatrix in the regime $\lambda\gtrsim2$. One can see that higher-order expansions yield significantly better accuracy.

\bibliography{reference}
\end{document}